\documentclass{article}
\usepackage{spconf,amsmath,graphicx}
\usepackage{color}
\usepackage {booktabs}
\usepackage{amssymb}
\usepackage{comment}
\usepackage{mathrsfs}
\usepackage{amsmath}
\usepackage{subcaption}
\usepackage{hyperref}
\usepackage{url}

\usepackage[
backend=biber,
style=ieee,
citestyle=numeric-comp,
maxbibnames=3,
maxcitenames=3,
doi=false,isbn=false,url=false,eprint=false
]{biblatex}

\usepackage{multirow}
\defbibheading{bibliography}[\refname]{}

\title{SingMOS: An extensive Open-Source Singing Voice Dataset \\ for MOS Prediction}
\name{Yuxun Tang$^{1}$, Jiatong Shi$^{2}$, Yuning Wu$^1$, Qin Jin$^{1}$*\thanks{*Corresponding author.}}
\address{
$^1$School of Information, Renmin University of China, P.R.China \\
  $^2$Language Technologies Institute, Carnegie Mellon University, U.S.A. \\
    \small{\texttt{\{tangyuxun, yuningwu, qjin\}@ruc.edu.cn, \{jiatongs\}@cs.cmu.edu}}
}
\begin{document}
\ninept
\maketitle 
\begin{abstract}
In speech generation tasks, human subjective ratings, usually referred to as the opinion score, are considered the "gold standard" for speech quality evaluation, with the mean opinion score (MOS) serving as the primary evaluation metric. Due to the high cost of human annotation, several MOS prediction systems have emerged in the speech domain, demonstrating good performance. These MOS prediction models are trained using annotations from previous speech-related challenges. However, compared to the speech domain, the singing domain faces data scarcity and stricter copyright protections, leading to a lack of high-quality MOS-annotated datasets for singing.
To address this, we propose SingMOS, a high-quality and diverse MOS dataset for singing, covering a range of Chinese and Japanese datasets. These synthesized vocals are generated using state-of-the-art models in singing synthesis, conversion, or resynthesis tasks and are rated by professional annotators alongside real vocals. Data analysis demonstrates the diversity and reliability of our dataset. Additionally, we conduct further exploration on SingMOS, providing insights for singing MOS prediction and guidance for the continued expansion of SingMOS.

\end{abstract}

\begin{keywords}
singing, evaluation, MOS, prediction
\end{keywords}

\section{Introduction}
\label{sec:intro}
Mean Opinion Score~(MOS) is considered as the "gold standard" for evaluating the quality of synthetic speech and singing clips, while other subjective metrics like Mel-cepstral distance~(MCD) show less correlation with perceptual audio quality~\cite{kubichek1993mel, vazquez2002reliability}, especially with recent high-fidelity synthesis models~\cite{wang2017tacotron, ren2020fastspeech, }.  
Typically, MOS relies on gathering subjective evaluations from listeners to quantify the perceived quality of audio samples. During MOS evaluations, a panel of listeners rates audio segments on a scale from 1 to 5, where 5 denotes excellent quality and 1 indicates poor quality. 
MOS allows researchers to gain direct insights into users' subjective perceptions of audio quality, providing crucial guidance for subsequent technological enhancements and optimizations.

However, it is generally a known issue in the community that MOS evaluations are time-consuming and expensive because of the need for successive amounts of effort from a range of listeners. Recently, pseudo MOS prediction models have been proposed to overcome the problems mentioned and achieved a high correlation to MOS scores in speech generation tasks. Many recent speech generation studies starts to use predicted MOS in their evaluation pipeline~\cite{ju2024naturalspeech, siuzdak2023vocos, sellam2023squid, chang2024dsu, jung2024espnet, wang2024usat}.

The very beginning of the success in automatic MOS neural predictor starts from MOSNet~\cite{Lo_2019}, which adopts convolutional neural networks and bidirectional long short-term memory networks for voice conversion evaluation.
Building on the foundation of MOSNet, MBNet~\cite{leng2021mbnet} advances it by using the opinion scores of each listener instead of the average of multiple scores. 
LDNet~\cite{huang2022ldnet}, based on MBNet, has been restructured to enhance its stability and efficiency.
As self-supervised learning~(SSL) models demonstrate astonishing capabilities in speech-related tasks, SSL models have also been applied to MOS prediction~\cite{tseng2021utilizing, cooper2022generalization,saeki22c_utmos, Tseng2022DDOSAM,yang2022fusion}.
Tseng et al.~\cite{tseng2021utilizing} propose a framework to facilitate pre-trained SSL models for MOS prediction.
Cooper et al.~\cite{cooper2022generalization} explore the generalization ability of MOS prediction networks and find that wav2vec2.0~\cite{baevski2020wav2vec} models fine-tuned for MOS prediction demonstrate the best results.
In VoiceMOS Challenge 2022~\cite{huang2022voicemos} and 2023~\cite{cooper2023voicemos}, numerous SSL-based models have emerged, significantly improving the accuracy of speech MOS prediction. Notable examples include UTMOS~\cite{saeki22c_utmos}, DDOS~\cite{Tseng2022DDOSAM}, and the fusion model~\cite{yang2022fusion} from VoiceMOS Challenge 2022, as well as LE-SSL-MOS~\cite{qi2023ssl} from VoiceMOS Challenge 2023.

Analyzing the datasets used by the aforementioned MOS prediction models reveals that all models are mostly trained and tested exclusively on speech data.
However, there are limited singing-related works, due to the scarcity of singing MOS data and stricter copyright protections.
Considering the domain gap between speech and singing, particularly in terms of higher requirements for pitch accuracy and naturalness in singing, it is challenging to directly apply speech MOS data to singing MOS prediction tasks~\cite{huang2023singing, cooper2023voicemos}.
Therefore, it is essential to construct a high-quality singing MOS dataset.

In this work, we propose SingMOS, the first publicly available high-quality dataset for singing MOS. 
SingMOS uses a subset of CtrSVDD~\cite{zang2024ctrsvdd}, including ground truth and synthetic vocals generated by state-of-the-art~(SOTA) models in singing synthesis, conversion, or resynthesis tasks, covering a range of Chinese and Japanese datasets.
To ensure diversity, SingMOS includes 21 types of singing voice synthesis~(SVS) models, 6 variants of singing voice conversion~(SVC) models, and 6 resynthesis~(vocoder) models to generate 3421 singing clips~(4.25 hours), with an average length of 4.47 seconds. 
All ground truth singing clips and corresponding music scores are from open-source singing datasets, ensuring an open-access benchmark.
Based on the dataset, we conduct additional explorations in singing MOS precition. Specifically, we present a baseline system over the SingMOS dataset and explore the effect of different SSL models and additional pitch-related information. 
The SingMOS dataset will be publicly accessible under CC-By-SA-NC 4.0 and the link will be provided soon.

\section{Dataset: SingMOS}
The SingMOS dataset comprises 3,421 Chinese and Japanese vocal clips with a sample rate of 16 kHz, totaling 4.25 hours in duration. In this section, we introduce the process of collecting the SingMOS dataset and describe the data split for training, development, and testing. Additionally, we provide a statistical analysis of the dataset.

\begin{figure}[ht]
    \centering
    \begin{subfigure}[b]{0.49\linewidth}
        \centering
        \includegraphics[width=\linewidth]{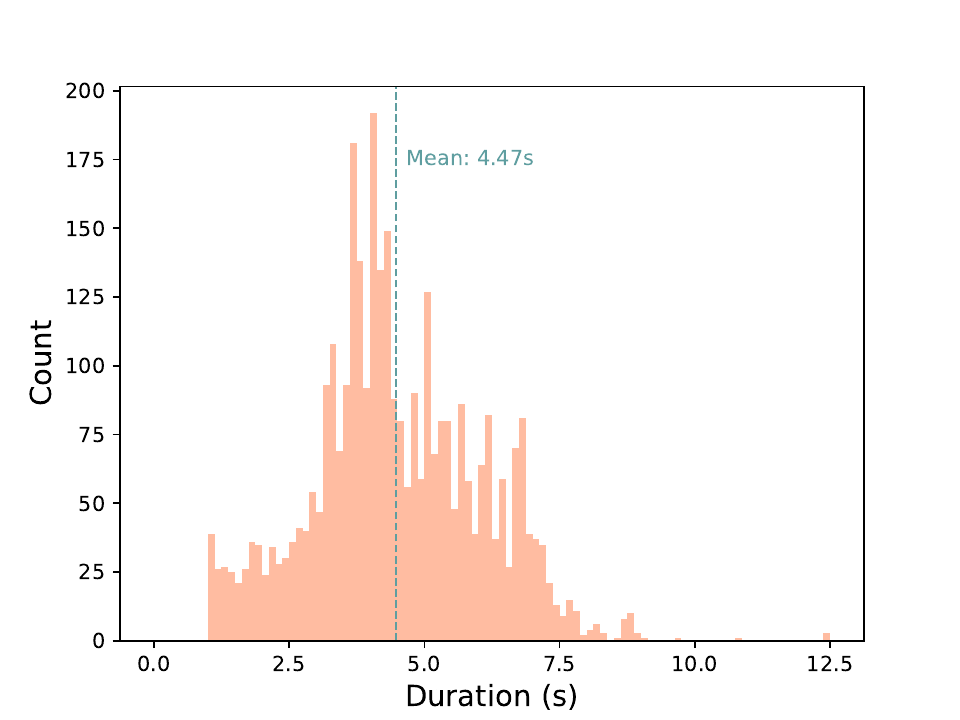}
        \caption{Overall Duration}
        \label{subfig:dur_overall}
    \end{subfigure}
    \hfill
    \begin{subfigure}[b]{0.49\linewidth}
        \centering
        \includegraphics[width=\linewidth]{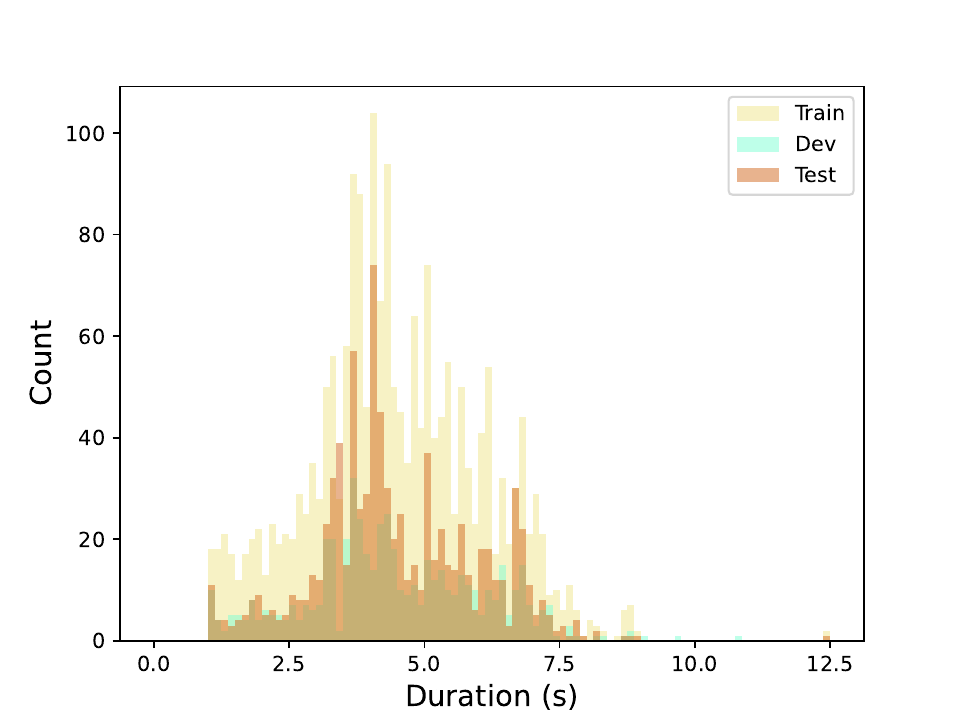}
        \caption{Sets Duration}
        \label{subfig:dur_sets}
    \end{subfigure}
    \\
    \begin{subfigure}[b]{0.49\linewidth}
        \centering
        \includegraphics[width=\linewidth]{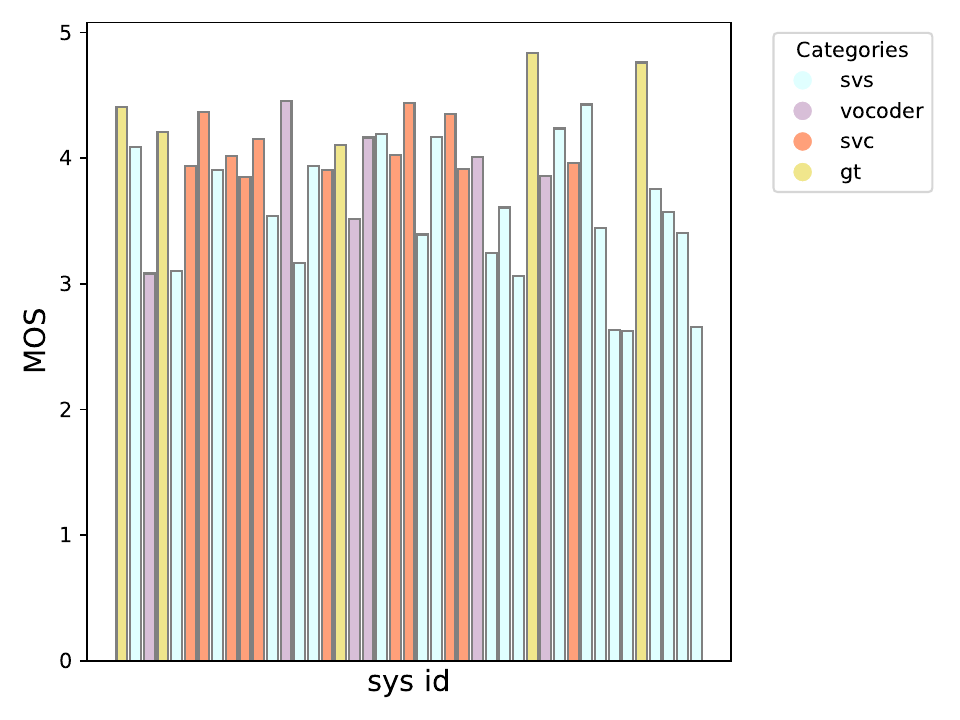}
        \caption{Overall System MOS}
        \label{subfig:sys_mos}
    \end{subfigure}
    \hfill
    \begin{subfigure}[b]{0.49\linewidth}
        \centering
        \includegraphics[width=\linewidth]{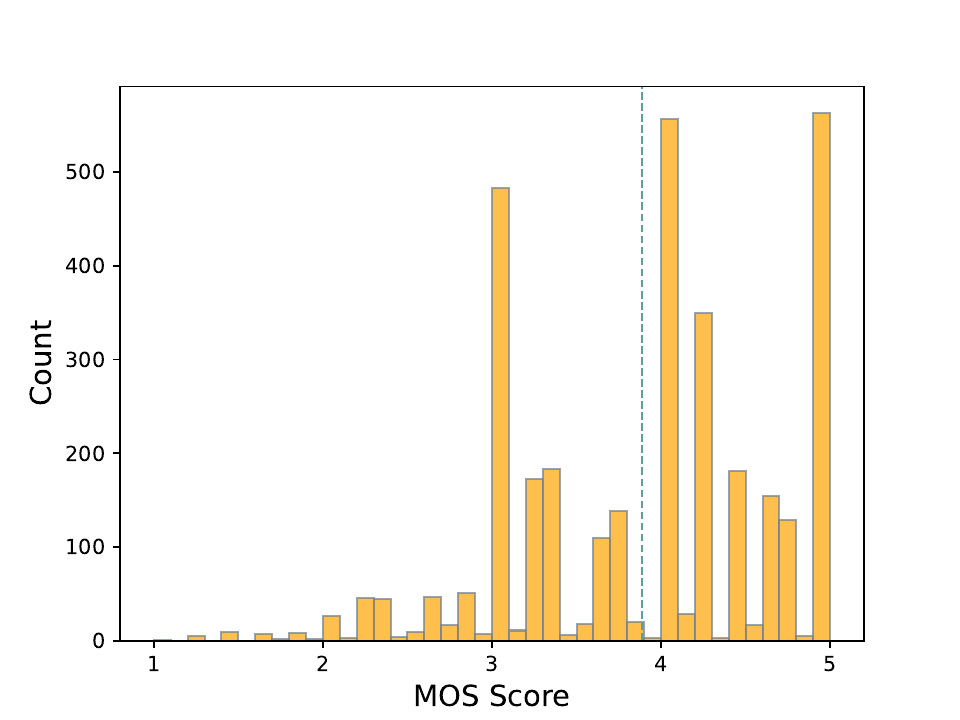}
        \caption{Overall Utterance MOS}
        \label{subfig:utt_mos}
    \end{subfigure}
    \caption{Histogram of the SingMOS dataset. The subfigure~\ref{subfig:dur_overall} shows the duration distribution of the whole dataset whereas the subfigure~\ref{subfig:dur_sets} demonstrates the duration distribution among the train/development/test splits. The subfigure~\ref{subfig:sys_mos} and the subfigure~\ref{subfig:utt_mos} show the overall system mos distribution and the utterance MOS distribution respectively. }
    \label{fig:1-dur&mos}
\end{figure}

\subsection{Data Collection}
All singing clips in SingMOS are prepared based on existing open source singing datasets, including Chinese singing datasets: Opencpop~\cite{wang2022opencpop}, M4Singer~\cite{zhang2022msinger}, ACE-opencpop~\cite{shi2024singing}, Kising~\cite{shi2024singing} and Japanese singing datasets: Kiritan~\cite{Ogawa2021E2074}, JVSMusic~\cite{tamaru2020jvsmusic}, Ofuton-P\footnote{\url{https://sites.google.com/view/oftn-utagoedb}}.
For all Chinese datasets, we adopt the official temporal segmentation provided in their original papers. 
For most Japanese datasets, we follow the process in ESPnet-Muskits toolkit~\cite{shi22muskits}, while systems from NNSVS follow the segmentation from its own split~\cite{yamamoto2023nnsvs}. 

To ensure the diversity of SingMOS, we follow the design in CtrSVDD~\cite{zang2024ctrsvdd} to select the \textbf{raw dataset} from the source singing datasets above, including ground truth data and synthetic data.

\noindent \textbf{Ground truth Data}:
We follow the train, development, and test split as defined in ESPnet for all Chinese and Japanese datasets~(the split setting in NNSVS is the same as ESPnet). Only test sets of datasets are used as ground truth data in the raw dataset. Due to the limited dataset size and license restrictions, we choose only five ground truth systems from ACE-Opencpop, Opencpop, M4Singer, and JVSMusic.

\noindent \textbf{Synthetic Data}:
We integrate 21 singing voice synthesis~(SVS) systems, 11 singing voice conversion~(SVC) systems, and six singing resynthesis~(vocoder) systems, ensuring both advanced capabilities and diversity. More systems details are shown in Sec~\ref{sec:sys-detail}. 
To ensure reproducibility, we primarily utilize models in open-source toolkits such as Muskits-ESPnet and NNSVS. All systems are trained in the corresponding dataset recipes of open-source toolkits, and we infer the trained system on the test set to obtain synthetic data in the raw dataset.

We filter out all singing segments shorter than one second in the raw dataset, as they are too short for effective subjective annotation.
Considering the impact of segmentation on some datasets, which resulted in limited singing samples, we do not consider systems with fewer than 50 samples in the train and development sets (the split setting in Sec.~\ref{sec:split_setting}).

\begin{figure}[ht]
    \centering
    \begin{subfigure}[b]{0.48\linewidth}
        \centering
        \includegraphics[width=\linewidth]{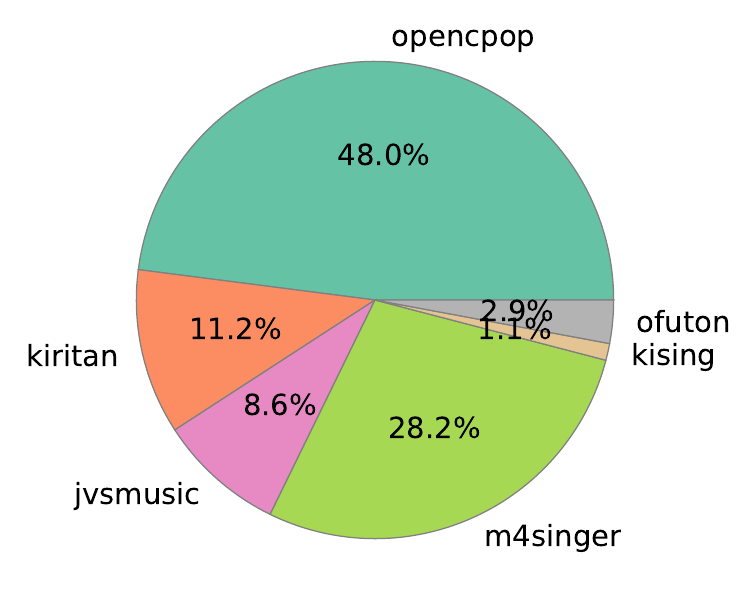}
        \caption{Source datasets on whole data}
        \label{subfig:dataset_overall}
    \end{subfigure}
    \begin{subfigure}[b]{0.48\linewidth}
        \centering
        \includegraphics[width=\linewidth]{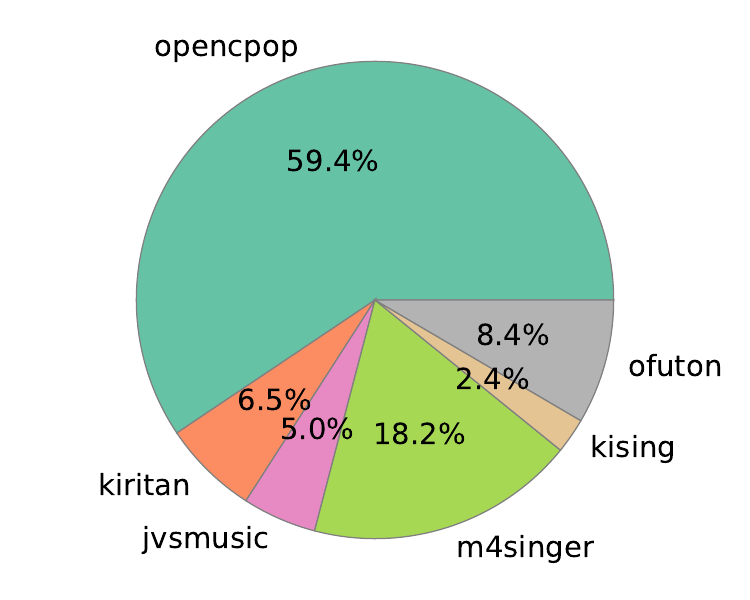}
        \caption{Source datasets on test sets}
        \label{subfig:dataset_test}
    \end{subfigure}
    \\
    \begin{subfigure}[b]{0.48\linewidth}
        \centering
        \includegraphics[width=\linewidth]{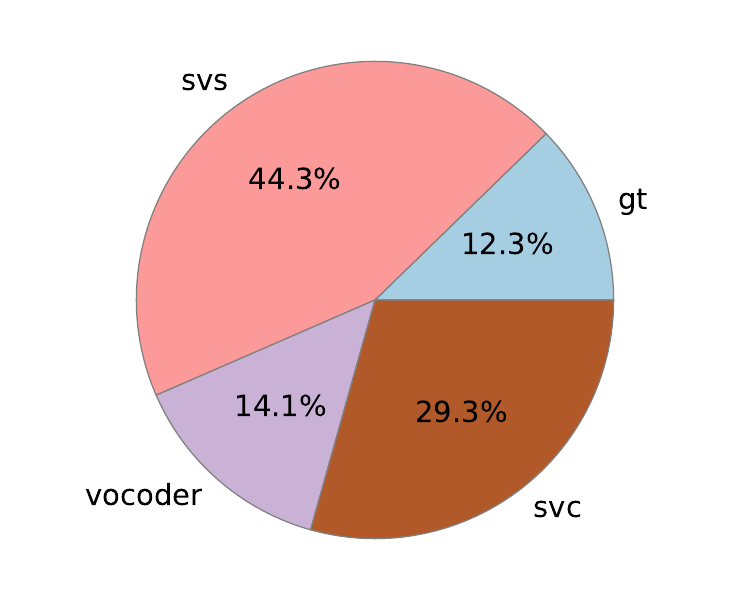}
        \caption{Models in systems of whole data}
        \label{subfig:sys_overall}
    \end{subfigure}
    \begin{subfigure}[b]{0.48\linewidth}
        \centering
        \includegraphics[width=\linewidth]{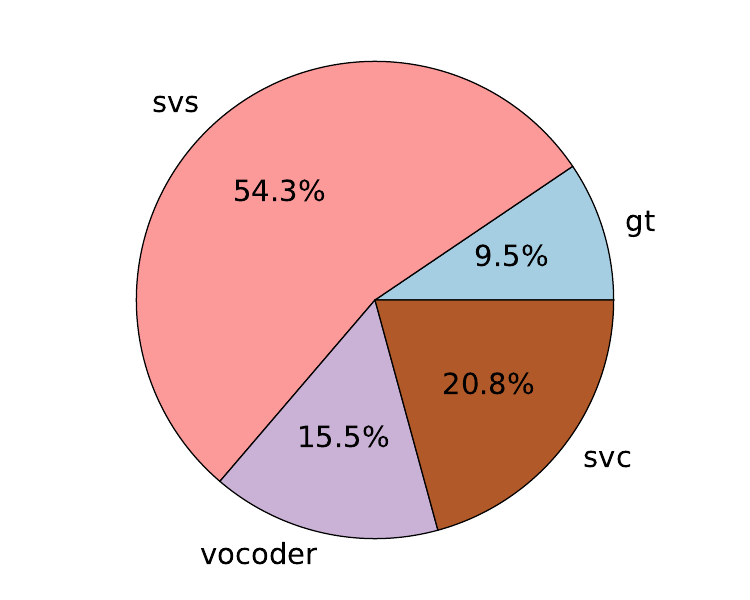}
        \caption{Models in systems of test sets}
        \label{subfig:sys_test}
    \end{subfigure}
    \caption{Overview of source datasets and models distribution on whole data and the test sets of SingMOS dataset.}
    \label{fig:2-sets}
\end{figure}

\subsection{Data Annotation}
To ensure the statistical significance of the subjective annotation, we target a sufficient number of annotations for each system in SingMOS.\footnote{We ensure that the multiplication of singing samples and listeners is approximately 500.}

While the majority of singing samples are from CtrSVDD dataset, we further balance the number of samples for each system from the original source. 
The chosen subset (i.e., main subset), comprises 31 systems, each with 90-100 singing samples, totaling 3000 singing samples. These samples are distributed to five listeners for annotation, obtaining around 15,000 MOS annotation points.
We also provide an unseen subset for held-out samples with four systems, each consisting of up to 50 samples. Each sample in the unseen subset is evaluated by at least ten annotators.
The unseen subset comprises a total of 189 singing samples and receives 1,890 MOS annotation points in total. During the annotation process, we further equally split the two subsets into ten batches, with five annotators for each batch.

To further enhance the diversity of SingMOS, we also utilize evaluation results from the Discrete Speech Challenge at Interspeech 2024~\cite{chang2024interspeech} and merge them into the SingMOS dataset. The additional samples contain 232 audio samples from seven SVS systems and ground truth data on Opencpop where each system composes 29 audio samples labeled by 20 listeners.

In general, the \textbf{SingMOS} dataset includes a total of 3,421 singing samples, with an average length of 4.47 seconds and a total of 21,530 MOS annotations. 
The distribution for song clips duration and MOS annotations is shown in Figure~\ref{fig:1-dur&mos} and an overview of source datasets and models in systems of whole data is illustrated in Figure~\ref{subfig:dataset_overall} and Figure~\ref{subfig:sys_overall}.

\begin{table}[t!]
    \centering
    \caption{Train-dev-test split of SingMOS dataset.}
    \label{tab:dataset}
    \centering
    \resizebox{\linewidth}{!}{
    \begin{tabular}{lcccc} 
        \toprule
        \toprule
        \textbf{Parition} & \textbf{\# Systems}  & \textbf{\# Utterances} & \textbf{\# Annotations}  & \textbf{Models}  \\
        \midrule
        Train & 31 & 2000 & 10000 & M01$\sim$M16 \\
        \midrule
        Development & 31+2 & 544 & 2940 & M01$\sim$M16 \\
        \midrule
        Test-main & 31+4 & 645 & 3950 & M01$\sim$M17 \\
        \midrule
        Test-other1 & 8 & 232 & 4640 & M07,M08,UKN \\  
        \bottomrule        
        \bottomrule
    \end{tabular}
    }
\end{table}

\subsection{Data Split}
\label{sec:split_setting}
To facilitate the application of SingMOS, we divide the dataset as follows:
For the main subset of SingMOS, we split the data into train, development, and test sets in a 60\%, 20\%, and 20\% ratio, respectively.
For the unseen subset (comprising a total of four systems), we split two systems (\textbf{M14} on Ofuton and \textbf{M12} on KiSing) equally between the development and test sets, and allocate the remaining two systems entirely to the test set~(\textbf{test-main}).
For the challenge evaluation results, we consider them as an additional test set~(\textbf{test-other1}).
After the division, four sets are obtained a train set, a development set, and two test sets. Table~\ref{tab:dataset} demonstrates detailed information.

\subsection{Details of Systems}
\label{sec:sys-detail}
Typically, a system refers to running a model in a dataset. Therefore, analyzing a system involves examining its model and dataset.
To provide a more detailed analysis of SingMOS, we conducted in-domain and out-domain analyses on systems of two test sets, including test-main and test-other1.

\noindent \textbf{In-Domain Data}: 
The in-domain data is used to evaluate whether the MOS prediction model has been effectively trained. 
Here it primarily comprises test-main except for unseen systems, which share the same systems as those in the train and development sets and consist of 20\% of the main subset in the raw dataset.
Datasets in the in-domain data mainly includes the following: M4Singer~\cite{zhang2022msinger}, Opencpop~\cite{wang2022opencpop}, Kiritan~\cite{Ogawa2021E2074}, JVSMusic~\cite{tamaru2020jvsmusic}, and ACE-Opencpop~\cite{shi2024singing}. 
Models in in-domain data are as follows:
\begin{itemize}
    \item SVS models: \textbf{M01} is Naive RNN~\cite{naivernn}; \textbf{M02} is XiaoiceSing~\cite{lu2020xiaoicesing}; \textbf{M03} is NNSVS~\cite{yamamoto2023nnsvs}; \textbf{M04} is DiffSinger~\cite{liu2022diffsinger}; \textbf{M05} is VISinger~\cite{VISinger}; \textbf{M06} is VISinger2~\cite{visinger2}; \textbf{M07} is TokSing~\cite{wu2024toksing}; \textbf{M08} is Discrete RNN with SingOMD tokens~\cite{tang2024singomd}.
    Most models appear in CtrSVDD~\cite{zang2024ctrsvdd} instead of \textbf{M07} and \textbf{M08}, two models with discrete tokens.
    \item SVC models: \textbf{M09} - \textbf{M12} are four variations of Soft-VITS-SVC, one of major frameworks adopted in the SVC challenge 2023~\cite{huang2023singing} with Chinese HuBERT~\cite{hsu2020hubert}, MR-HuBERT~\cite{shi2024multiresolution}, WavLM~\cite{chen2022wavlm} and ContentVec~\cite{qian2022contentvec}. \textbf{M13} is the Nagoya University (NU) SVC model~\cite{yamamoto2023nusvc}. More details of these models can be found in CtrSVDD~\cite{zang2024ctrsvdd}.
    \item vocoder models: \textbf{M14} is HiFi-GAN~\cite{kong2020hifigan} and \textbf{M15} is unit-HiFiGAN, a viration for discrete tokens. \textbf{M16} is based on official ACE-Studio~\cite{shi2024singing}.
\end{itemize}
 
\noindent \textbf{Out-Domain Data}: 
The out-domain data is utilized to test the generalization ability of the MOS prediction model. Systems generating out-domain data are not present in the SingMOS train set.
Due to the limited number of systems in two sets, we perform analyses on unseen datasets and models for each set respectively.


\begin{itemize}
    \item Test-main: The SVS model \textbf{M05} on the Ofuton-P dataset, the SVC model \textbf{M12} on the Kising dataset~\cite{shi2024singing}, and the vocoder \textbf{M14} on the Ofuton-P dataset are considered as unseen dataset systems. \textbf{M17}, a vocoder model, which is a variation of unit-HiFiGAN with SingOMD tokens~\cite{tang2024singomd}, running on the Opencpop dataset, is considered as an unseen model.
    \item Test-other1: Models in the set are all run on the Opencpop dataset and consist of three variations of \textbf{M07}, one \textbf{M08}, and three unknown models~(\textbf{UKN}). All systems. All systems, except for one based on \textbf{M07} and one based on \textbf{M08}, are considered as unseen systems.
\end{itemize}


\section{Singing MOS Prediction}
Singing MOS prediction is an application of MOS prediction in the domain of singing, bearing certain similarities to speech MOS prediction. 
However, due to the higher demands for pitch accuracy and naturalness, and the need to handle more diverse scenarios in singing compared to speech, singing MOS prediction is a more challenging task. 

\subsection{Baseline and Experiment Setting}
In speech MOS prediction, \cite{tseng2021utilizing,cooper2022generalization} explore the approach of fine-tuning speech SSL models for MOS prediction, achieving promising results. 
Considering the capabilities and generalization of speech SSL models, we adopt the approach of fine-tuning SSL models as our baseline and further explore models better suited for singing MOS prediction.

Specifically, we design the baseline by referring to the method implemented\footnote{\url{https://github.com/nii-yamagishilab/mos-finetune-ssl}} in \cite{cooper2022generalization}: fine-tune an SSL model by mean-pooling the model’s output embedding features, adding a linear output layer, and training with L1 loss. 

To measure the performance of the MOS predictor, we follow prior work~\cite{tseng2021utilizing, cooper2022generalization} and utilize mean squared error (MSE) along with various correlation metrics that reveal whether the relative orderings of the scores are predicted correctly. The Linear Correlation Coefficient (LCC) is usually considered a basic correlation measure. The Spearman Rank Correlation Coefficient (SRCC) is non-parametric and measures the correlation of ranking order, while the Kendall Tau Rank Correlation (KTAU) is a rank-based correlation that tends to be more robust to errors.

For fine-tuning settings, we fine-tune the model on the train set of SingMOS for 50 epochs using one NVIDIA 3090Ti, selecting the best model based on the validation set. 

\begin{figure*}[t]
    \centering
    \begin{subfigure}[b]{0.18\linewidth}
        \centering
        \includegraphics[width=\linewidth]{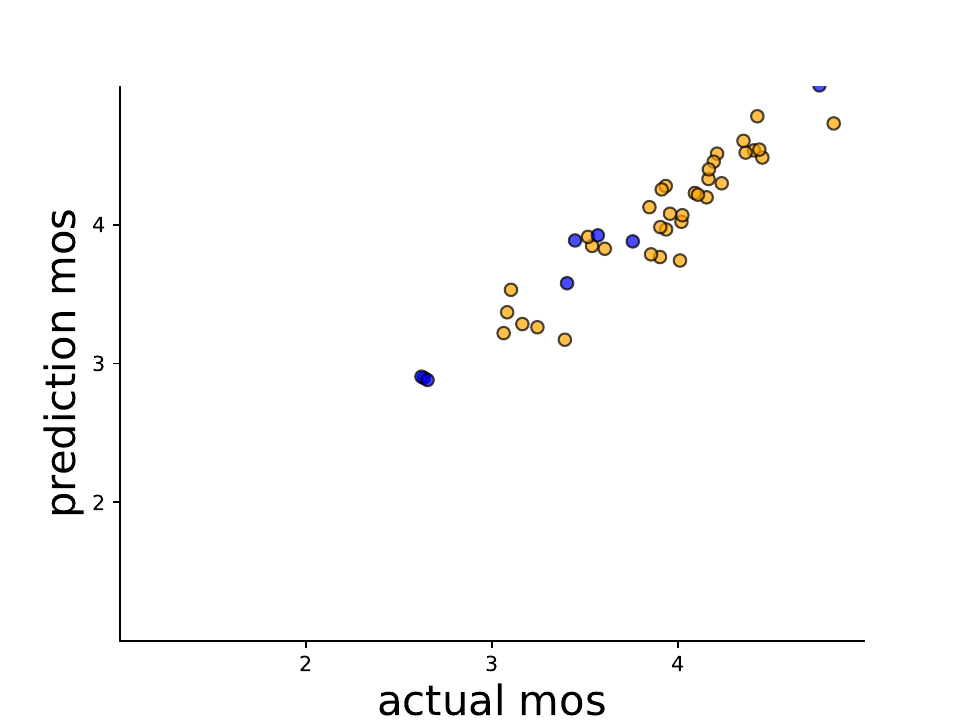}
        \caption{wav2vec2.0}
    \end{subfigure}
    \begin{subfigure}[b]{0.18\linewidth}
        \centering
        \includegraphics[width=\linewidth]{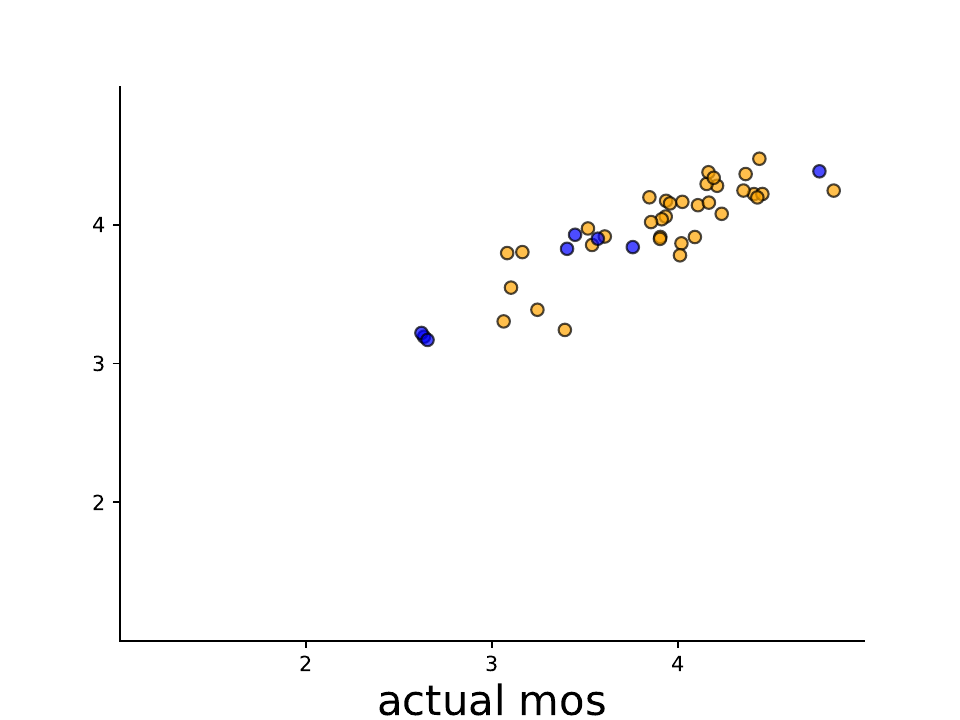}
        \caption{wav2vec2.0*}
    \end{subfigure}
    \begin{subfigure}[b]{0.18\linewidth}
        \centering
        \includegraphics[width=\linewidth]{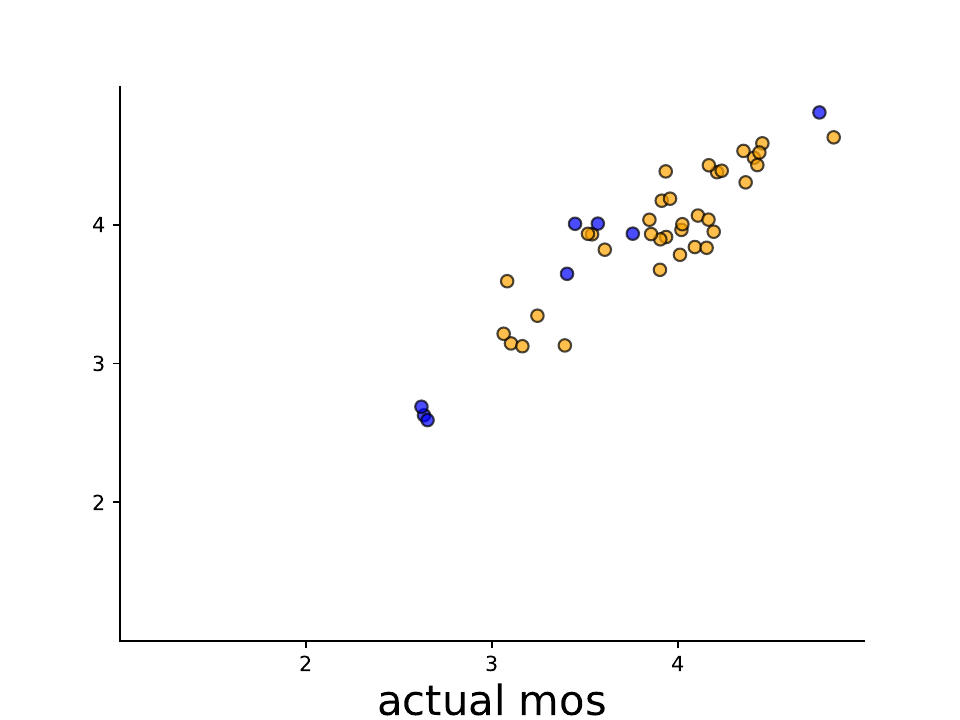}
        \caption{HuBERT}
    \end{subfigure}
    \begin{subfigure}[b]{0.18\linewidth}
        \centering
        \includegraphics[width=\linewidth]{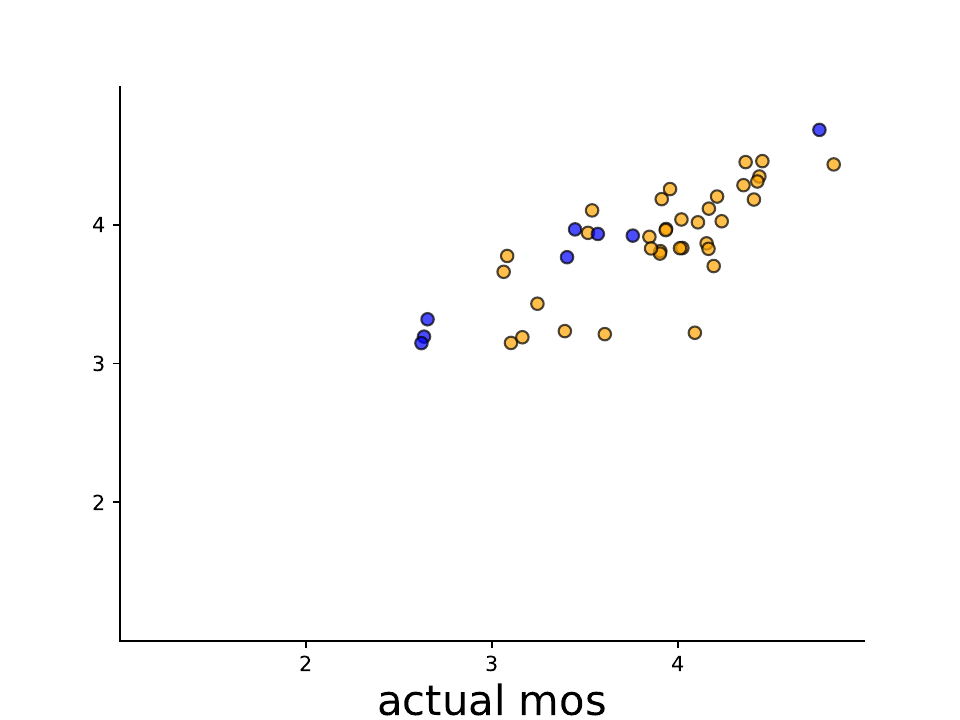}
        \caption{HuBERT*}
    \end{subfigure}
    \begin{subfigure}[b]{0.18\linewidth}
        \centering
        \includegraphics[width=\linewidth]{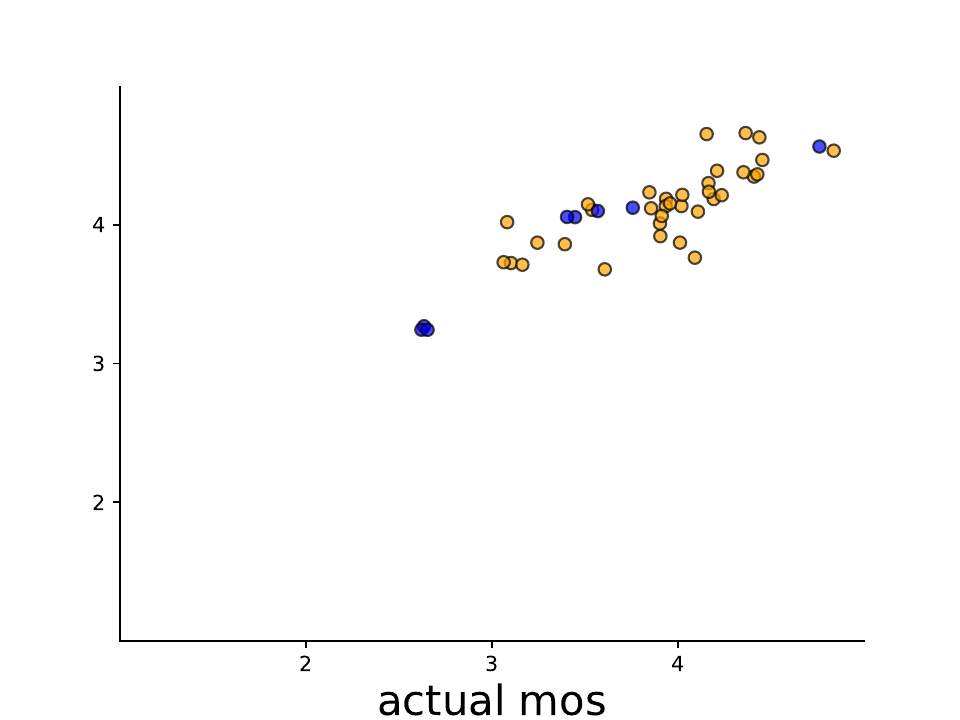}
        \caption{XLS-R*}
    \end{subfigure}
    \begin{subfigure}[t]{0.07\linewidth}
        \centering
        \includegraphics[width=\linewidth, height=2cm]{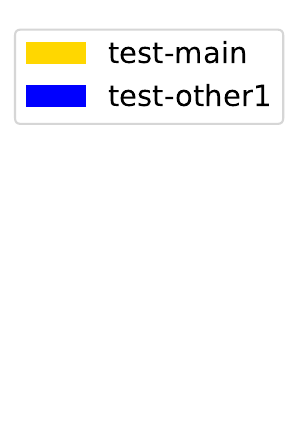}
    \end{subfigure}
    \caption{Scatter plot of system-level prediction results for each SSL model. * indicates large models with over 300M parameters.}
    \label{fig:3-mos}
\end{figure*}

\subsection{Exploration on SSL models}
\label{sec:ssl}

\begin{table}[h]
    \centering
    \caption{Results of different SSL models on test-main.  * indicates large models with over 300M parameters.}
    \label{tab:test-main}
    \centering
    \resizebox{\linewidth}{!}{
        \begin{tabular}{l|cccc|cccc} 
            \toprule
            \multirow{2}{*}{\textbf{SSL Model}} & \multicolumn{4}{c|}{\textbf{Utterance}} & 
                \multicolumn{4}{c}{\textbf{System}}  \\
            & \textbf{Test Error$\downarrow$}  & \textbf{LCC$\uparrow$} & \textbf{SRCC$\uparrow$}  & \textbf{KTRCC$\uparrow$} & 
                \textbf{Test Error$\downarrow$}  & \textbf{LCC$\uparrow$} & \textbf{SRCC$\uparrow$}  & \textbf{KTRCC$\uparrow$} \\
            \midrule
            wav2vec2.0 & \textbf{0.381} & \textbf{0.665} & \textbf{0.660} & \textbf{0.490} & 
                \textbf{0.046 }& \textbf{0.926} & \textbf{0.921} & \textbf{0.754} \\
            wav2vec2.0* & 0.403 & 0.595 & 0.586 & 0.429 &
                0.075 & 0.810 & 0.812 & 0.610 \\
            HuBERT & 0.386 & 0.622 & 0.616 & 0.452 & 
                0.048 & 0.878 & 0.843 & 0.660 \\
            HuBERT* & 0.543 & 0.447 & 0.459 & 0.324 &
                0.094 & 0.721 & 0.738 & 0.566 \\
            XLS-R* & 0.507 & 0.498 & 0.504 & 0.364 &
                0.128 & 0.771 & 0.815 & 0.616 \\
            \bottomrule        
        \end{tabular}
    }
\end{table}

\begin{table}[h]
    \centering
    \caption{Results of different SSL models on test-other1.  * indicates large models with over 300M parameters.}
    \label{tab:test-other1}
    \centering
    \resizebox{\linewidth}{!}{
        \begin{tabular}{l|cccc|cccc} 
            \toprule
            \multirow{2}{*}{\textbf{SSL Model}} & \multicolumn{4}{c|}{\textbf{Utterance}} & 
                \multicolumn{4}{c}{\textbf{System}}  \\
            & \textbf{Test Error$\downarrow$}  & \textbf{LCC$\uparrow$} & \textbf{SRCC$\uparrow$}  & \textbf{KTRCC$\uparrow$} & 
                \textbf{Test Error$\downarrow$}  & \textbf{LCC$\uparrow$} & \textbf{SRCC$\uparrow$}  & \textbf{KTRCC$\uparrow$} \\
            \midrule
            wav2vec2.0 & 0.270 & \textbf{0.861} & 0.817 & 0.620 & 
                0.078 & \textbf{0.990} & 0.833 & 0.642 \\
            wav2vec2.0* & 0.423 & 0.777 & 0.802 & 0.598 &
                0.201 & 0.964 & 0.809 & 0.571 \\
            HuBERT & \textbf{0.246} & \textbf{0.861} & \textbf{0.826} & \textbf{0.631} &
                \textbf{0.076} & 0.965 & 0.833 & 0.642 \\
            HuBERT* & 0.430 & 0.805 & 0.769 & 0.567 &
                0.201 & 0.985 & \textbf{0.904} & \textbf{0.785} \\  
            XLS-R* & 0.586 & 0.736 & 0.725 & 0.520 & 
                0.299 & 0.956 & \textbf{0.904} & \textbf{0.785} \\
            \bottomrule        
        \end{tabular}
    }
\end{table}

In speech MOS prediction, \cite{cooper2022generalization} explores the generalization capabilities of various SSL models and finds that the wav2vec2.0-base model achieves the best results. 
However, due to the domain gap between singing and speech, directly applying conclusions from speech to singing may not be appropriate. 
Therefore, we further investigate the effectiveness of different pre-trained speech SSL models specifically in the context of singing MOS prediction.

We conduct experiments using various SSL models including HuBERT \cite{hsu2020hubert}, wav2vec 2.0 \cite{baevski2020wav2vec}, and XLS-R \cite{babu2021xls} from the Fairseq toolkit \footnote{\url{https://github.com/pytorch/fairseq}}, which have demonstrated robust capabilities in various speech tasks and shown good performance in MOS prediction as referenced in \cite{cooper2022generalization}.

Table~\ref{tab:test-main} and Table~\ref{tab:test-other1} show the performance of different speech SSL models on two test sets, respectively. 
In Table~\ref{tab:test-main}, the results for different SSL models on test-main indicate that wav2vec2.0-base achieves the best performance, outperforming other SSL models across all metrics. This is consistent with findings in~\cite{cooper2022generalization} on speech MOS prediction. It suggests that although current SSL models are primarily designed for speech-related tasks, they demonstrate good generalization capabilities and can effectively understand singing samples.
Table~\ref{tab:test-other1} presents the results of different SSL models on test-other1. The experimental findings indicate that, at the utterance level, the HuBERT-base achieves the best performance, with the wav2vec2.0-base closely matching its results. At the system level, HuBERT-base and wav2vec2.0-base show comparable performance, while HuBERT-large achieves the best results in terms of SRCC and KTRCC. It is worth noting that test-other1 mostly comprises unseen systems with relatively few utterances and systems in the entire set, yet several SSL models still demonstrate strong performance.

Figure~\ref{fig:3-mos} visualizes the relationship between predicted system MOS and true system MOS for different SSL models. Models closer to the diagonal from bottom-left to top-right indicate more accurate predictions. The figure shows that wav2vec2.0-base and HuBERT-base closely follow the diagonal on both test sets, which is consistent with the experimental results.

\subsection{Exploration on Additional Information}
In the VoiceMOS Challenge 2022~\cite{huang2022voicemos} and Voice MOS Challenge 2023~\cite{cooper2023voicemos}, various models~\cite{huang2022ldnet, saeki22c_utmos} begin incorporating additional information to improve prediction accuracy, such as using listener IDs to utilize each annotator's ratings and phoneme sequences recognized by speech recognition models.
Considering that pitch accuracy and the naturalness of pitch variation are important aspects of MOS annotation for singing, we introduce F0 information and F0 variance into the model to investigate their effects.

According to Sec~\ref{sec:ssl}, we choose wav2vec2.0-base as the baseline model and conduct a comparison on test-main. To incorporate F0 and F0 variance, we add an additional embedding layer respectively, and concatenate the embedding feature with SSL embedding features. The follow-up parts are the same as the baseline with a mean-pooling and a linear layer.
The experimental results are shown in the table~\ref{tab:additional}.

\begin{table}[h]
    \centering
    \caption{Results of adding additional information on test-main}
    \label{tab:additional}
    \centering
    \resizebox{\linewidth}{!}{
        \begin{tabular}{l|cccc|cccc} 
            \toprule
            \multirow{2}{*}{\textbf{Method}} & \multicolumn{4}{c|}{\textbf{Utterance}} & 
                \multicolumn{4}{c}{\textbf{System}}  \\
            & \textbf{Test Error$\downarrow$}  & \textbf{LCC$\uparrow$} & \textbf{SRCC$\uparrow$}  & \textbf{KTRCC$\uparrow$} & 
                \textbf{Test Error$\downarrow$}  & \textbf{LCC$\uparrow$} & \textbf{SRCC$\uparrow$}  & \textbf{KTRCC$\uparrow$} \\
            \midrule
            Baseline & \textbf{0.381} &\textbf{ 0.665} & \textbf{0.660} & \textbf{0.490} & 
                0.046 & \textbf{0.926} & \textbf{0.921} & \textbf{0.754} \\
            Baseline+F0 & 0.392 & 0.635 & 0.634 & 0.466 & 
                \textbf{0.041} & 0.923 & 0.891 & 0.710 \\
            Baseline+F0 var. & 0.444 & 0.620 & 0.611 & 0.449 & 
                0.084 & 0.923 & 0.877 & 0.700 \\
            \bottomrule        
        \end{tabular}
    }
\end{table}

Experimental results show that adding f0 and f0 variance as auxiliary information does not improve Singing MOS Prediction and even worsens the results. One possible reason is that our current dataset is limited in its scale for the model to fully utilize the additional f0-related information, leading to poorer performance. Another potential reason is that speech SSL models already have some capability to extract F0-related information from the audio~\cite{lin2023utility}. In future work, we plan to expand the SingMOS dataset and conduct further investigations.

\section{Conclusion}
In this paper, to address the lack of publicly available MOS datasets in the singing domain, we firstly propose a public available dataset, SingMOS, which covers various Chinese and Japanese datasets as well as state-of-the-art models in singing tasks. We analyze the dataset from multiple perspectives to validate its reliability and diversity. Furthermore, we build a baseline for Singing MOS Prediction based on our dataset and conduct a further exploration, providing a reference for future advancements.

\vfill\pagebreak
\newpage

\section{References}
{
\printbibliography
}
\end{document}